\documentstyle[11pt,aaspp4]{article}
\newcommand{\postscript}[2]
 {\setlength{\epsfxsize}{#2\hsize}
   \centerline{\epsfbox{#1}}}

\newcommand{\btheta}{\mbox{\boldmath $\theta$}}
\newcommand{\bdelta}{\mbox{\boldmath $\delta$}}
\newcommand{\bzeta}{\mbox{\boldmath $\zeta$}}
\newcommand{\bz}{\mbox{\boldmath $z$}}

\begin{document}

\title{On the Astrometric Behavior of Binary Microlensing Events}
\author
{Cheongho Han}
\affil{Department of Astronomy \& Space Science, Chungbuk National University,
Chongju, Korea 361-763; cheongho@astro.chungbuk.ac.kr}

\begin{abstract}
Despite the suspected binarity for a significant fraction of Galactic lenses, 
the current photometric surveys detected binary microlensing events only for 
a small fraction of the total events.  The detection efficiency is especially 
low for non-caustic crossing events, which comprise majority of the binary 
lensing events, due to the absence of distinctive features in their light 
curves combined with small deviations from the standard light curve of a
single point-mass event.  In addition, even they are detected, it will be 
difficult to determine the solution of the binary lens parameters due to the 
severe degeneracy problem.  In this paper, we investigate the properties of 
binary lensing events expected when they are astrometrically observed by using 
high precision interferometers.  For this, we construct vector field maps of 
excess centroid shifts, which represent the deviations of the binary lensing 
centroid shifts from those of a single lensing events as a function of source 
position.  From the analysis of the maps, we find that the excess centroid 
shifts are substantial in a considerably large area around caustics.  In 
addition, they have characteristic sizes and directions depending strongly 
on the source positions with respect to the caustics and the resulting 
trajectories of the light centroid (astrometric trajectories) have distinctive 
features, which can be distinguished from the deviations caused by other 
reasons.  We classify the types of the deviations and investigate where they 
occur.  Due to the strong dependency of the centroid shifts on the lens system 
geometry combined with the distinctive features in the observed astrometric 
trajectories, astrometric binary lensing observations will provide an 
important tool that can probe the properties of Galactic binary lens 
population.

\end{abstract}

\section{Introduction}

Searches for Galactic dark matter by monitoring light variations of stars 
located in the Magellanic Clouds and Galactic bulge caused by gravitational 
microlensing have been or are being carried out by several groups (MACHO: 
Alcock et al.\ 2000; EROS: Afonso et al.\ 1999; OGLE: Udalski, Kubiak, \& 
Szyma\'nski 1997).  The searches have been very successful and the total 
number of candidate events now reaches up to 1,000 (P.\ Popowski 2000, 
private communication).  Among these events, a notable fraction of events 
are believed to be caused by binary lenses  (Udalski et al.\ 1994; Bennett 
et al.\ 1995; Alard, Mao, \& Guibert 1995; Rhie \& Bennett 1996; Pratt et 
al.\ 1996; Albrow et al.\ 1996; Dominik \& Hirshfeld 1996; Alcock et al.\ 
1999a, 1999b).  Detecting binary lensing events is important because one 
can obtain valuable information about Galactic binary lens population such 
as the distributions of binary mass ratios and separations.

Due to the difference in the lens system geometry, the light curve of a 
binary lensing event deviates from the single-peak symmetric one of an
event caused by a single point-mass lens.  The most important difference 
in the geometry of a binary lens system from that of a single lens system 
is the formation of caustics.  When a source star crosses the caustics, 
an extra pair of images appear (or disappear) and its flux is greatly 
amplified, resulting characteristic sharp spikes in the light curve.  
If the source does not cross the caustics, on the other hand, the light 
curve is less perturbed and in most cases has no distinctive features such 
as the spikes of the caustic crossing event.  Di Stefano \& Perna (1997) 
pointed out that the detection efficiency of non-caustic crossing binary 
lensing events is significantly lower than that of caustic crossing events 
due to the absence of distinctive features in their light curves combined 
with small deviations.  This was well demonstrated by the recent MACHO 
result where non-caustic crossing events comprise only $\sim 25\%$ of the 
total detected binary lensing events despite their predicted higher frequency
of occurrence (Alcock et al.\ 1999b).  Another problem in investigating 
binaries from microlensing is that even if an event is detected, it is 
difficult to determine the solution of the binary lens parameters, especially 
for non-caustic crossing events, due to the severe degeneracy problem (Mao 
\& Di Stefano 1995).  The level of degeneracy is so high for most gently 
perturbed light curves of non-caustic crossing events that many of these 
events allow fits to models of the binary companion ranging from equal-mass 
binary to planets (Di Stefano \& Perna 1997).  Although the level of 
degeneracy for caustic crossing events is relatively low due to the 
characteristic spike feature in their light curves, they cannot be totally 
free from the degeneracy problem (Dominik 1999; Albrow et al.\ 1999).  
Therefore, for the acquisition of accurate information about Galactic 
binaries from microlensing, it will be essential to devise a new 
method that can effectively detect binary lensing including non-caustic 
crossing events and can resolve their degeneracy problem.

In addition to photometry, microlensing can also be detected and observed 
astrometrically.  When a source star is lensed by a single point-mass lens, 
its image is split into two.  The typical separation between the two images 
for a Galactic event caused by a stellar mass lens is on the order of a 
milli-arcsec.  This separation is too small for direct observation of the 
individual images even with the {\it Hubble Space Telescope}.  However, if 
microlensing is observed by using several planned high-precision 
interferometers from space-based platform, e.g.\ the {\it Space Interferometry 
Mission} (SIM, http://sim.jpl.nasa.gov), and ground-based large telescopes, 
e.g.\ the Keck and the Very Large Telescope, one can measure the 
astrometric displacements in the source star image light centroid, 
$\bdelta\btheta_{\rm c}$.  Once the centroid shifts are measured and their 
trajectory is constructed, one can determine the lens proper motion, from 
which one can better constrain the lens mass compared to the photometrically 
obtained Einstein ring radius crossing time $t_{\rm E}$.  Astrometric 
microlensing observation was first proposed nearly simultaneously by 
Miyamoto \& Yoshii (1995), H\o\hskip-1pt g, Novikov, \& Polnarev (1995), 
and Walker (1995) and further analysis considering actual performances of 
specific instrument has been done by Boden, Shao, \& Van Buren (1998).  A 
modified idea of monitoring a nearby high proper motion star working as a 
lens instead of monitoring a source star for the accurate determination of 
the lens mass was proposed by Miralda-Escud\'e (1996) and Paczy\'nski (1998).
For a single lensing event, the light centroid traces out an elliptical 
trajectory during the event (Walker 1995; Jeong, Han, \& Park 1999).  If 
an event is caused by a binary, on the other hand, the resulting trajectory 
of the light centroid (astrometric trajectory) deviates from the elliptical 
one of the single lensing event (Chang \& Han 1999), which is analogous 
to the deviations in the photometric light curve.

In this paper, we investigate the properties of binary lensing events when 
they are astrometrically observed by using high precision interferometers.  
For this, we construct vector field maps of excess centroid shifts, 
$\Delta\bdelta\btheta_{c}(\xi,\eta)$, which represent the deviations of the 
binary lensing centroid shifts from those of a single lensing event as a 
function of source position $\bzeta=(\xi,\eta)$.  From the analysis of the 
maps, we find that the excess centroid shifts are substantial in a 
considerably large area around caustics.  In addition, they have 
characteristic sizes and directions depending strongly on the source positions 
with respect to the caustics and the resulting astrometric trajectories have 
distinctive features which can be distinguished from the deviations caused 
by other reasons.  We classify the types of the deviations and investigate 
where they occur.

\section{Basics of Binary Lensing}

When lengths are normalized to the combined Einstein ring radius, the lens 
equation in complex notations for a binary lens system is represented by
$$
\zeta = z + {m_{1} \over \bar{z}_{1}-\bar{z}} 
+ {m_{2} \over \bar{z}_{2}-\bar{z}},
\eqno(1)
$$
where $m_1$ and $m_2=1-m_1$ are the mass fractions of the individual 
lenses, $z_1$ and $z_2$ are the positions of the lenses, $\zeta=\xi+i\eta$ 
and $z=x+iy$ are the positions of the source and images, and $\bar{z}$ 
denotes the complex conjugate of $z$ (Witt 1990).  The combined Einstein 
ring radius is related to the total mass and the location of binary the 
lens by
$$
r_{\rm E} = \left( {4GM\over c^2} 
{D_{ol}D_{ls}\over D_{os}}\right)^{1/2},
\eqno(2)
$$
where $D_{ol}$, $D_{ls}$, and $D_{os}$ represent the separations between 
the observer, lens, and source star, respectively.  For a binary lensing
event, the number of images is either three or five depending on the 
source position.  The amplifications of the individual images are given 
by the Jacobian of the transformation (1) evaluated at the image position, 
i.e.\
$$
A_i = \left({1\over \vert {\rm det}\ J\vert} \right)_{z=z_i};
\qquad {\rm det}\ J = 1-{\partial\zeta\over\partial\bar{z}}
{\overline{\partial\zeta}\over\partial\bar{z}}.
\eqno(3)
$$
The caustic refers to the source position on which the amplification of 
a point source event becomes infinity, i.e.\ ${\rm det}\ J=0$.  The total 
amplification is given by the sum of the amplifications of the individual 
images, i.e.\ $A=\sum_i A_i$.  Since the position of the image centroid 
equals to the amplification weighted mean position of the individual images, 
the source image centroid shift with respect to the unlensed source position 
is given by
$$
\bdelta\btheta_{\rm c} = {\sum_i A_i \bz_i \over \sum_i A_i} - \bzeta,
\eqno(4)
$$
where the positions of the individual images $\bz_i$ are obtained by 
numerically solving the lens equation (1).

\section{Properties of Various Types of Binary Lensing Events}

The size of caustics depends strongly on the binary separation $\ell$ 
(normalized by the combined angular Einstein ring radius $\theta_{\rm E}
=r_{\rm E} /D_{ol}$), causing strong dependency of the caustic crossing 
probability on the binary separation.  Caustic crossing is optimized when 
the binary sepration is equivalent to the angular Einstein ring radius.  
Han (1999) estimated that when the binary separation is in the range 
$0.7\lesssim \ell \lesssim 1.8$, the caustic crossing probability is greater 
than $\sim 50\%.$\footnote{He estimated the probability under the definition 
of a binary lensing event as `a close lens-source encounter within the 
combined Einstein ring radius with its center at the center of mass of the 
binary lens system'.}  However, as the separation further increases or 
decreases, the caustic crossing probability decreases rapidly.  Therefore, 
most non-caustic crossing events are caused by binaries with separations 
either substantially smaller (narrow binaries) or larger (wide binaries) 
than $\theta_{\rm E}$.  In this work, we focus on the astrometric properties 
of events caused by these types of binary lenses because they are much 
more common.

\subsection{Narrow Binary Lensing Events}

If an event is caused by a narrow binary, the lens system forms 3 
separated closed caustics.  One of them is located near the center of mass 
(the central caustic) and has a diamond shape with 4 cusps.  The other 
two are located away from the center of mass (the outer caustics) and 
each of them has a wedge-like shape with 3 cusps.  The outer caustics 
are symmetric with respect to the binary axis and are located on the 
heavier lens side with respect to the center of mass.  Compared to the 
central caustic, the outer caustics are much smaller (see  Figure 1).
If the binary separation is very small, both the central and outer 
caustics become very small and the geometry of the lens system mimics 
that of a single lensing event with a mass equal to the total mass of the 
binary and located at the center of mass of the binary (Gaudi \& Gould 
1997).  For a narrow binary lensing event, we, therefore, define the 
{\it excess centroid shift} by
$$
\Delta\bdelta\btheta_{\rm c} = 
\bdelta\btheta_{\rm c,b} - \bdelta\btheta_{\rm c,s},
\eqno(5)
$$
where $\bdelta\btheta_{\rm c,b}$ and $\bdelta\btheta_{\rm c,s}$ represent 
the centroid shifts of the binary and the corresponding single lensing 
event, respectively.

In Figure 1, we present the vector field map of the excess centroid 
shifts, $\Delta\bdelta\btheta_{\rm c}(\xi,\eta)$, for an example narrow 
binary lens system with $\ell=0.5$ and $q=0.2$.  The positions in the map 
are chosen so that the center of mass of the binary is at the origin.  
Both lenses are located on the $\xi$ axis and the heavier lens is to the 
right.  The circle drawn with a long-dashed line represents the combined 
Einstein ring and the closed figures drawn by thick solid line are caustics.
Also drawn on the map are the contours of the absolute value of the excess 
centroid shift, $\Delta\delta\theta_{\rm c}$.  The contours are drawn at 
the levels of $\Delta\delta\theta_{\rm c}=0.02\theta_{\rm E}$ (dotted line) 
and $0.2\theta_{\rm E}$ (solid line).  The angular Einstein ring radius 
of a Galactic bulge event caused by a stellar mass lens with $M\sim 0.3
\ M_\odot$ located at the half way between the observer and the source, 
i.e.\ $D_{ol}/D_{os}=0.5$, is $\theta_{\rm E}\sim 300$ $\mu$-arcsec.  
Therefore, these thresholds correspond to $\Delta\delta\theta_{\rm c}\sim 6$ 
$\mu$-arcsec and $\Delta\delta\theta_{\rm c}\sim 60$ $\mu$-arcsec, which are 
respectively related to the astrometric precisions of the SIM (Unwin et al.\ 
1997) and the interferometers on the large ground-based telescopes (Colavita 
et al.\ 1998; Mariotti et al.\ 1998).  To better show $\Delta\bdelta
\btheta_{\rm c} (\xi,\eta)$ near the caustic regions, we expand map of the 
regions and presented in Figure 2: upper panel for the central caustic 
region and lower panel for the outer caustic region.  In Figure 3, we also
present the light curves (upper 6 panels) and the astrometric trajectories
(lower 6 panels) for several example non-caustic crossing events whose 
source trajectories are marked in Figure 1a (straight lines with arrows).
Note that the number in each panel of Fig.\ 3 matches with the number of 
the corresponding source trajectory in Fig.\ 1.

From the analysis of $\Delta\bdelta\btheta_{\rm c}(\xi,\eta)$ in various 
regions outside the caustics\footnote{
Since our major interest is the astrometric properties of non-caustic 
crossing events, in this section we limit our discussion only about excess 
centroid shifts in the regions outside of caustics.  The properties of 
$\Delta\bdelta\btheta_{\rm c}$ in the region inside of caustics will be 
discussed in \S\ 3.3
} 
and the resulting light 
curves and astrometric trajectories, one finds the following patterns of 
the excess centroid shifts.  First, as already known by Chang \& Han (1999), 
the amount of excess centroid shifts is substantial in a considerably large 
area around caustics, c.f.\ the excess amplification map of Gaudi \& Gould 
(1997), implying that astrometric binary lensing observation is an efficient 
method in detecting non-caustic crossing events caused by narrow binary 
lenses.\footnote{
It would be useful to estimate the average amount of deviation expected 
from a typical Galactic event.  For this estimation, it is required to know 
not only the physical distributions of Galactic binaries and their mass 
function but also the distributions of their mass ratios and separations.
However, our knowledge about these quantities are very poor and thus we 
just compare the expected deviation for events caused by an example binary 
lens system.
}
The excess centroid shifts are especially large in the regions along the 
binary axis and the lines connecting the central and outer caustics.  In 
the region along the binary axis, the excess centroid shifts diverges from 
the on-binary-axis cusps of the central caustic.  Then, if a source passes 
through this region (e.g.\ the source trajectories 4 and 5), the resulting 
astrometric trajectory becomes {\it convex} compared to the elliptical one 
of the corresponding single lensing event.  By contrast, in the regions 
along the lines connecting the central and outer caustics, the excess 
centroid shifts converge towards the off-binary-axis cusps of the central 
caustic.  Then, the astrometric trajectory of an event with a source 
trajectory passing through this region (e.g.\ the source trajectories 1, 
2, and 3) will become {\it concave} with respect to that of the corresponding 
single lensing event or even {\it twisted}.  Due to these various types of 
deviations, the astrometric trajectories of non-caustic crossing binary 
lensing events have distinctive feaures.  By contrast, the light curves 
in most cases have no strong characteristic features.

\subsection{Wide Binary Lensing Events}

If an event is caused by a wide binary, on the other hand, the lens system
forms 2 separate closed caustics, which are located close to the individual 
lenses along the binary axis.  Each caustic has a wedge-like shape with 4 
cusps (see Figure 4).  For an event caused by these systems, the observed 
light curve and the astrometric trajectory appear to be the superpositions 
of those resulting from two events in which the component lenses behave as 
if they are two independent single lenses (Di Stefano \& Mao 1996; Di 
Stefano \& Scalzo 1999; An, Han, \& Park 2000).  Therefore, we define the 
excess centroid shifts of a wide binary lensing event by
$$
\Delta\bdelta\btheta_{\rm c} = 
\bdelta\btheta_{\rm c,b} - \bdelta\btheta_{\rm c,s,comb},
\eqno(6)
$$
where the combined centroid shift $\bdelta\btheta_{\rm c,s,comb}$ represents 
the vector sum of the centroid shifts caused by the individual single lenses.

In Figure 4, we present excess centroid shift map for an example wide binary 
lens system with $\ell=2.0$ and $q=0.5$.  In the figure, the two circles in 
the map drawn with a long-dashed line represent the Einstein rings of the 
individual single lenses.  Besides that there are two Einstein rings, the 
positions and notations are same as those in Fig.\ 1.  Maps in the regions 
around caustics are presented in Figure 5.  In Figure 6, we also present the 
light curves and astrometric trajectories for several example events resulting 
from the source trajectories marked in Fig.\ 4.

From the figures, one finds that $\Delta\bdelta\btheta_{\rm c}$ is also 
large for the wide binary lens system in a considerably large area around 
caustics and the resulting astrometric trajectories have distinctive 
features.  Especially large excess centroid shifts occur in the region 
along the binary axis.  One also finds that the general patterns of 
$\Delta\bdelta\btheta_{\rm c}$ in the region around each caustic is similar 
to those in the region around the central caustic of the narrow binary 
lens case.  That is, the excess centroid shifts diverges from the 
on-bianry-axis cusps and converges towards the off-binary-axis cusps.  
One major difference is that in the region between the right-side 
on-binary-axis cusp of the left caustic and left-side on-binary-axis cusp 
of the right caustic, the combination of the two diverging excess centroid 
shift vectors from the individual cusps makes $\Delta\bdelta\btheta_{\rm c}$ 
directed almost vertically from the binary axis.  As a result, the centroid 
shift trajectory of an event passing through this region (e.g.\ part of 
the trajectory 3 around the moment of the source star's closest approach 
to the center of mass) becomes convex compared to the astrometric trajectory 
of the combined single lensing events.  Another interesting deviations occur 
when the source passes closely both lenses.  For this case, the astrometric 
trajectory will have double bumps or twisted loops (e.g.\ the trajectories 
1, 2, and 3), which characterize the event is caused by a wide binary.

\subsection{Caustic Crossing Evenst}

When binary separation becomes equvalent to $\theta_{\rm E}$, the separated 
caustics merge together and form a single large closed caustic encompassing 
the center of mass of the binary.  The caustic has 6 cusps and two of them 
are located on the binary axis (see Figure 7).  Since the size of the caustic 
is maximized at around this binary separation, events caused by these binaries 
have high chance of caustic crossings.

To see the astrometric properties of caustic crossing events, we present 
in Figure 7 the vector field map of a binary lens system with $\ell=1.0$ 
and $q=0.5$.  We also present in Figure 8 two example astrometric 
trajectories (lower two panels) whose source trajectories are marked in 
the map.  Although this binary lens system cannot be well approximated 
either by a single lens nor by the superposition of the two independent 
single lenses, we compute the excess centroid shifts by using equation (5).  
From the figures, one finds the general pattern of the deviation vector in 
the region inside of the caustic is that $\Delta\bdelta\btheta_{\rm c}$ 
converges towards the on-binary-axis cusps and diverges from the 
off-binary-axis cusps.  Note that this pattern is exactly opposite to that 
of the excess centroid shifts in the region outside of the caustic.  As a 
result, the two adjacent excess centroid shift vectors just inside and outside 
of the caustic line have opposite directions.  When a source crosses the 
caustic, then, the deviations in the astrometric trajectory becomes very big 
not only because the absolute value of the centroid shift is very large near 
the caustics but also the direction of $\Delta\bdelta\btheta_{\rm c}$ is 
reversed during the caustic crossing.  We note that these patterns of 
$\Delta\bdelta\btheta_{\rm c}$ in the region very close to and inside of 
caustics apply also to caustic regions of the narow and wide binary lens 
systems (see the lower panels of Figure 1 and 3).  The only exception is 
the outer caustic of the narrow binary lens system, in which the direction 
of $\Delta\bdelta\btheta_{\rm c}$ is not reversed during caustic crossings 
(see the lower right panel of Fig.\ 2).

\section{Confusion with Other Types of Deviations}

In previous section, we illustrate various forms of deviations in the 
astrometric trajectories caused by the lens binarity.  Besides the lens 
binarity, however, the trajectory can also be distorted due to various other 
reasons.  First, the astrometric trajectory can be distorted by the change of 
the observer's location during the event caused by the orbital motion of the 
Earth around the Sun: parallax effect (Paczy\'nski 1998; Han 2000).  Second, 
if a star not participating in the lensing process is located too close to 
the source star (not the lens) to be resolved even with a high resolution 
interferometer, such as a close companion of the source star, the astrometric 
trajectory can also be distorted: blending effect (Han \& Kim 1999).  Third, 
if the size of the source star is not negligible compared to the Einstein 
ring radius, the astrometric trajectory is distorted: finite source effect 
(Mao \& Witt 998).

However, the distortions caused by these effects have different forms from 
those affected by the lens binarity.  The trajectory affected by the blending 
effect has characteristic linear component caused by the displacement towards 
the blended star [see the example trajectories in Figure 4 of Han \& Kim 
(1999)], and thus can be distinguished from the trajectories affected by the 
lens binarity.  The deviation caused by the parallax effect is periodic with 
a known period of a year, allowing one to notice the effect.  The devations 
caused by the finite source effect is somewhat confusing because it can 
produce concave and twisting distortions similar to those of binary lensing 
events.  However, the resulting astrometric trajectories of events affetced 
by the finite source effect are exactly symmetric with respect to the 
semi-minor axis of the unperturbed elliptical trajectory.  On the other hand, 
the astrometric trajectories of binary lensing events are asymmetric in 
general, and thus one can distinguish the two types of astrometric deviations.

\section{Summary}

The findings about the astrometric properties of binary lensing events from 
the analysis of the vector field maps of excess centroid shifts for various 
types of binary lens systems are summarized as follows.
\begin{enumerate}
\item
The lens binarity causes large deviations in the astrometric trajectory 
from the elliptical one of a single lensing event.  The excess centroid shifts 
are substantial even in the regions considerably away from the caustics, 
implying that the distortions will be significant even for a significant 
fraction of non-caustic crossing binary lensing events.

\item
Depending on the locations with respect to lens caustics, the excess centroid 
shifts have very characteristic sizes and directions.  As a source passes 
through these regions, the resulting astrometric trajectories have distinctive 
features, which are categorized by concave, convex, and twisting distortions.

\item
The binary-lens induced deviations in the astrometric trajectory can be 
distinguished from those caused by other effects, such as the parallax, 
blending, and finite source effects. 
\end{enumerate}
Therefore, astrometric microlensing observations will be an efficient method 
to detect events caused by binary lenses and to constrain their parameters 
with accuracy.

{\bf Acknowledgements}: 
This work was supported by grant KRF-99-041-D00442 of the Korea Research
Foundation.

\clearpage

\postscript{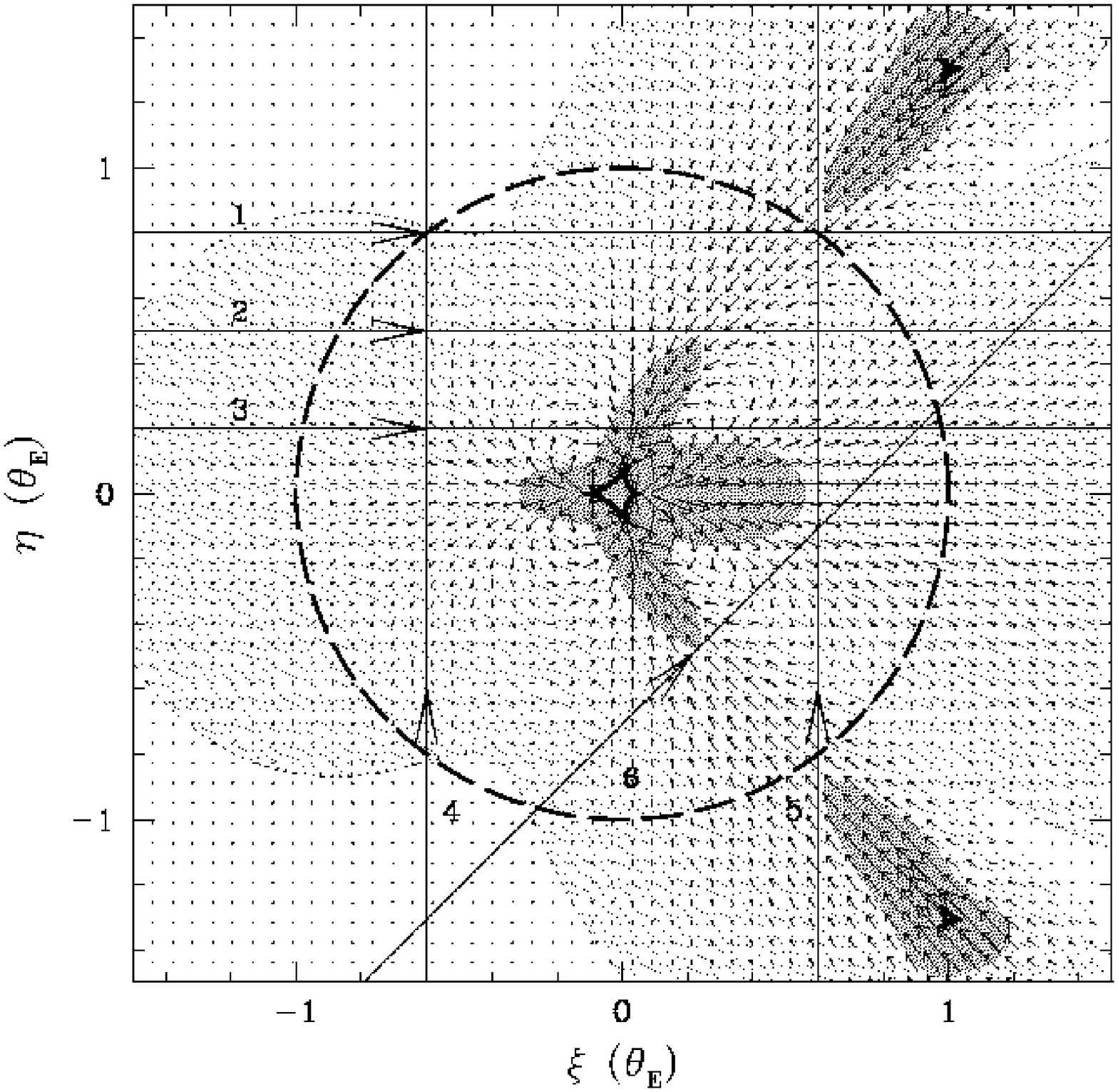}{1.0}
\noindent
{{\bf Figure 1:}\
Vector field map of the excess centroid shifts for an example narrow binary 
lens system with a separation $\ell=0.5$ and a mass ratio $q=0.2$.  The 
positions are chosen so that the center of mass is at the origin.  Both 
lenses are located on the $\xi$ axis and the heavier lens is to the right.  
The circle drawn with a long-dashed line represents the combined Einstein 
ring.  The three closed figures drawn by thick line are caustics.  Also 
drawn are the contours of the absolute values of the excess centroid shift, 
$\Delta\delta\theta_{\rm c}$.  The contours are drawn at the levels of 
$\Delta\delta\theta_{\rm c}=0.02 \theta_{\rm E}$ (dotted line) and $0.2
\theta_{\rm E}$ (solid line).  The straight lines with arrows represent the 
source trajectories whose resulting light curves and astrometric trajectories 
are presented in Fig.\ 3.  
}

\postscript{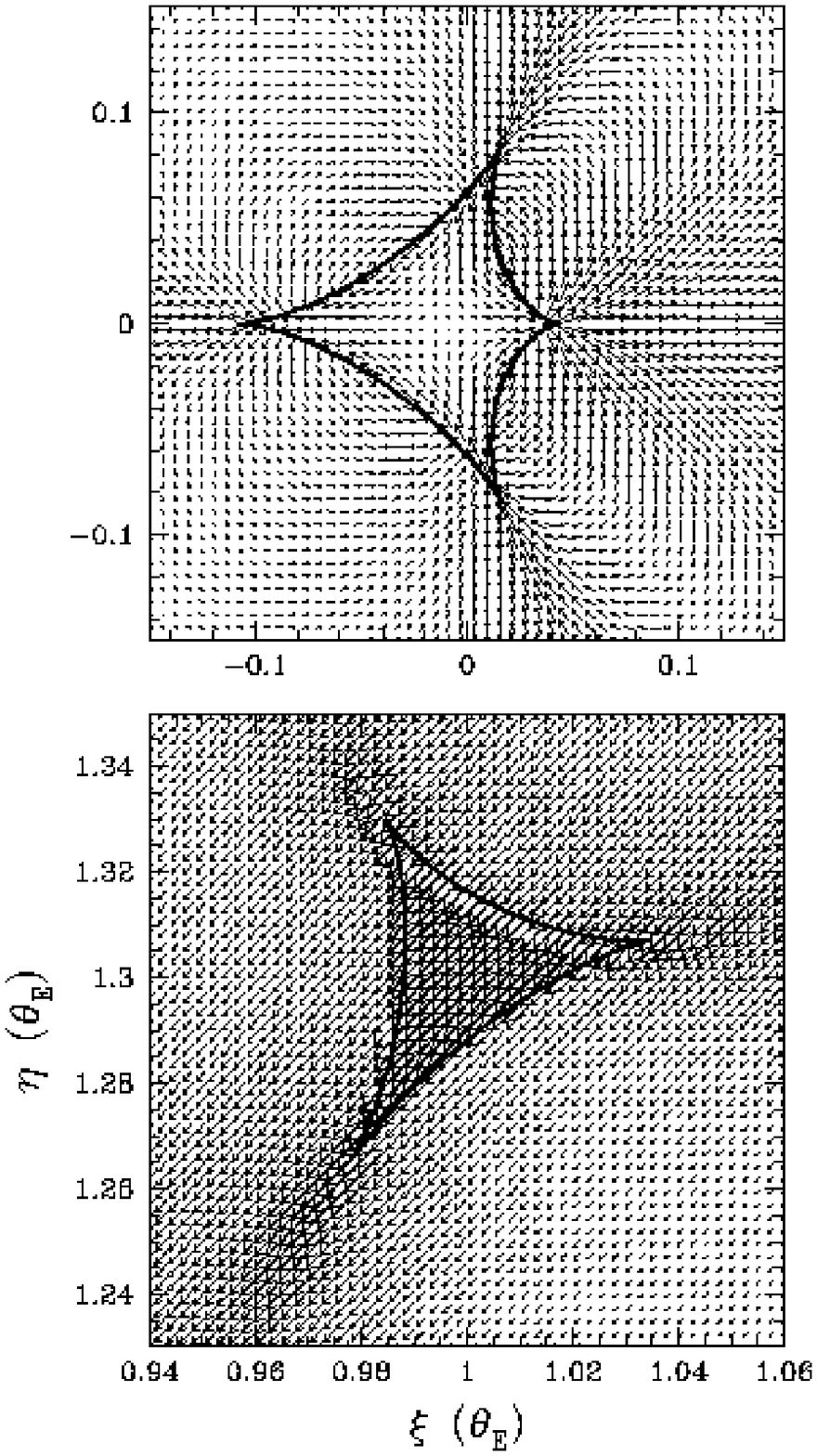}{1.2}
\noindent
{{\bf Figure 2:}\
Vector field map of the excess centroid shifts in the region near the caustics 
of the same lens system as in Fig.\ 1.  The upper and the lower panels are for 
the central and the outer caustic regions, respectively
}

\postscript{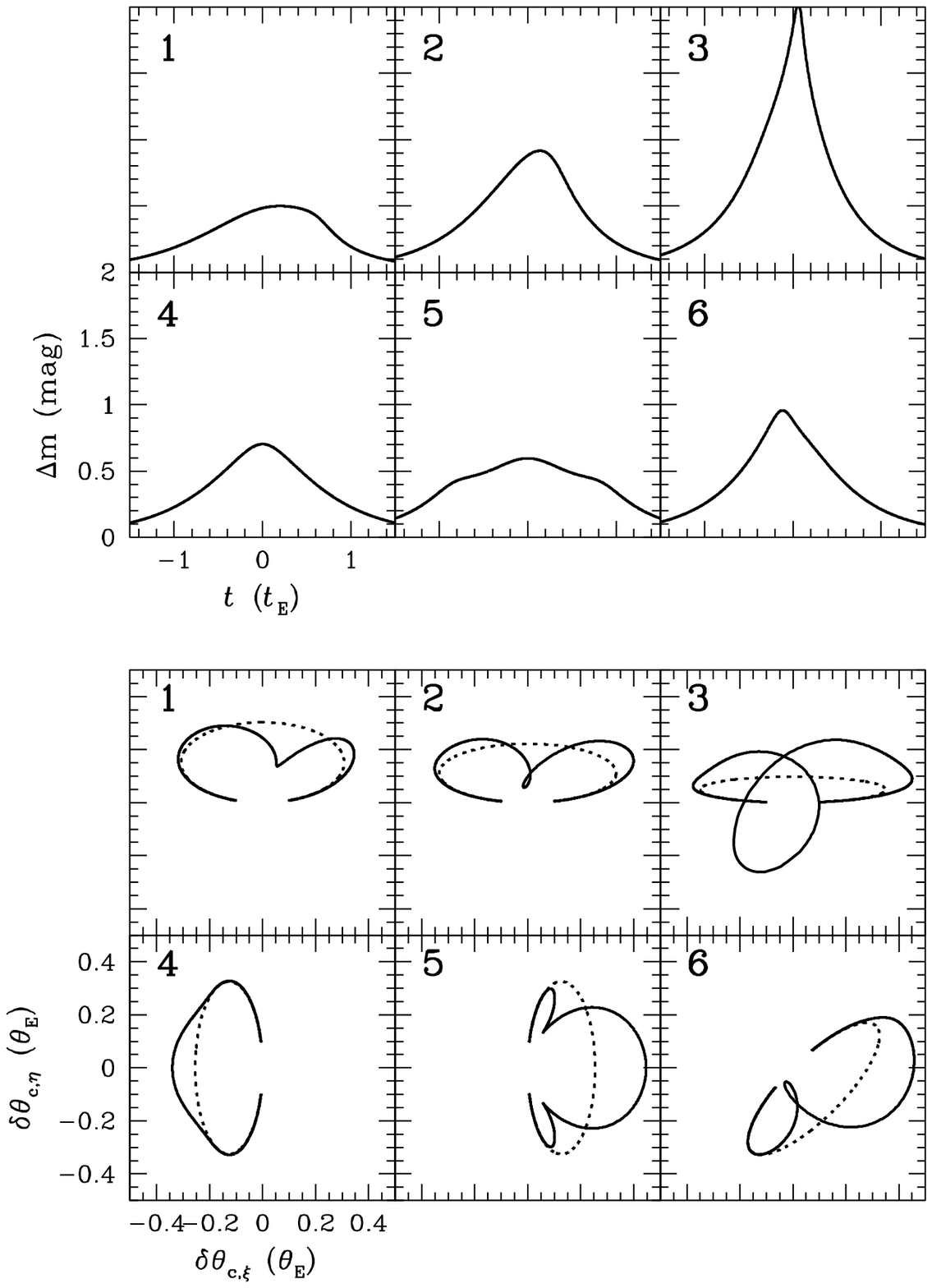}{1.15}
\noindent
{{\bf Figure 3:}\
Light curves (upper 6 panels) and astrometric trajectories (lower 6 panels) 
for several example non-caustic crossing narrow binary lensing events whose
source trajectories are marked in Fig\ 1.  The number in each panel matches 
with the number of the corresponding source trajectory in Fig.\ 1.  The 
dotted curve in each of the lower set of panels represents the astrometric 
trajectory of a single lensing event with a mass equal to the total mass of 
the binary and located at the center of mass of the binary.
}

\postscript{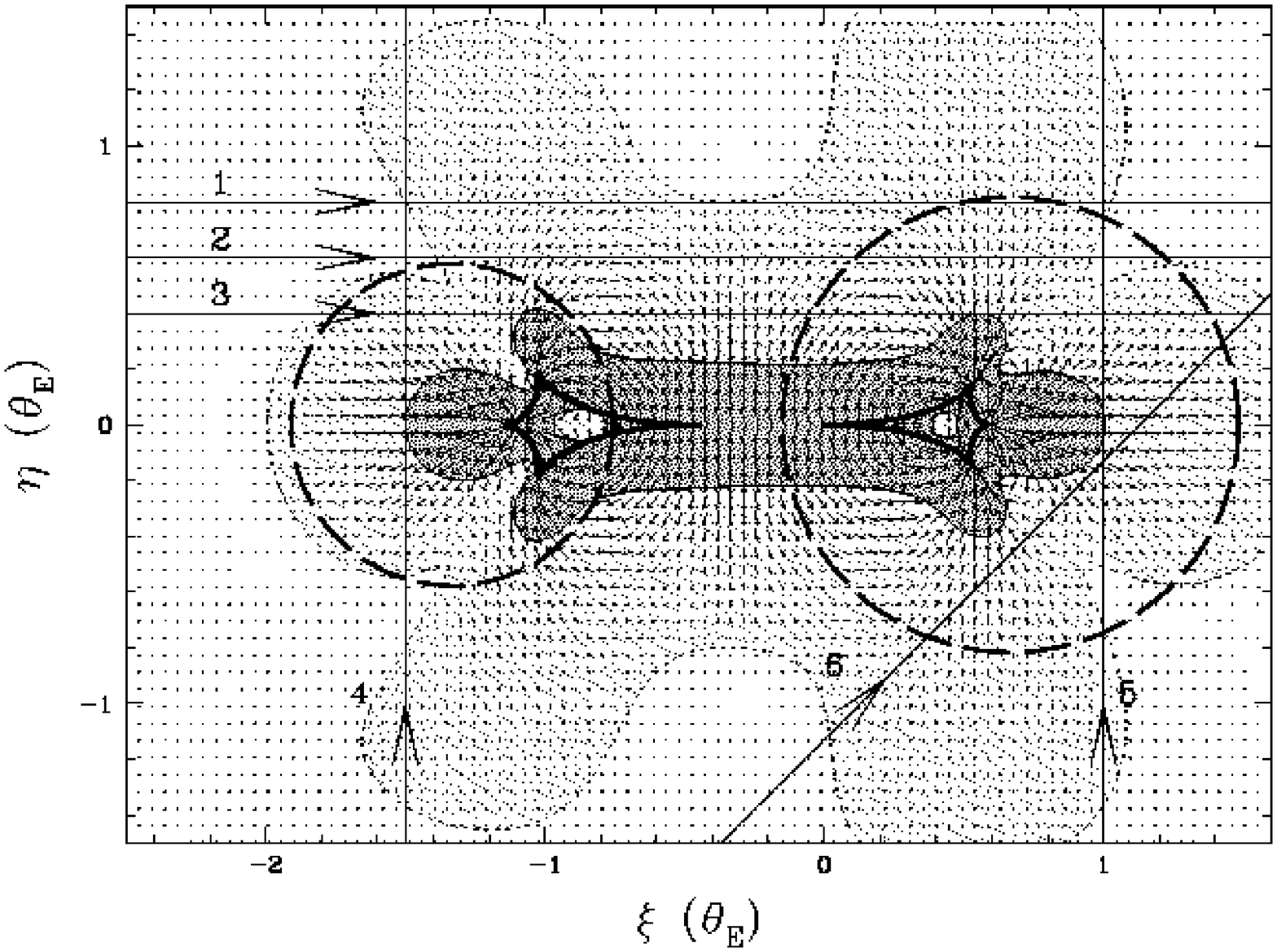}{1.0}
\noindent
{{\bf Figure 4:}\
Vector field map of the excess centroid shifts for an example wide binary 
lens system with a separation $\ell=2.0$ and a mass ratio $q=0.5$.  The two 
circles drawn by a long-dashed line represent the Einstein rings of the 
individual single lenses.  The positions and notations are same as those 
in Fig.\ 1.
}

\postscript{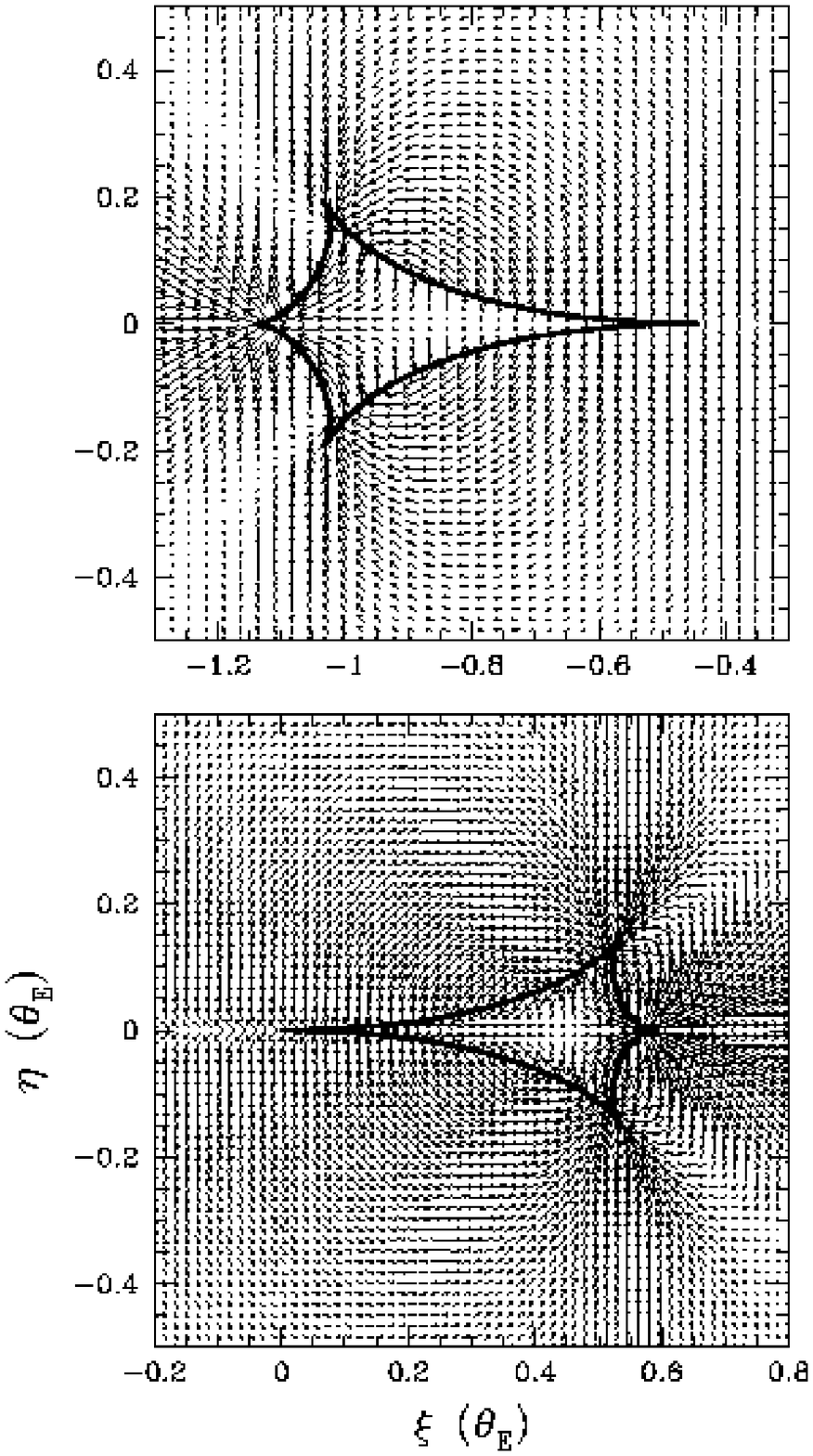}{1.2}
\noindent
{{\bf Figure 5:}\
Vector field map of the excess centroid shifts in the region near the 
caustics of the same lens system as in Fig.\ 4.
}

\postscript{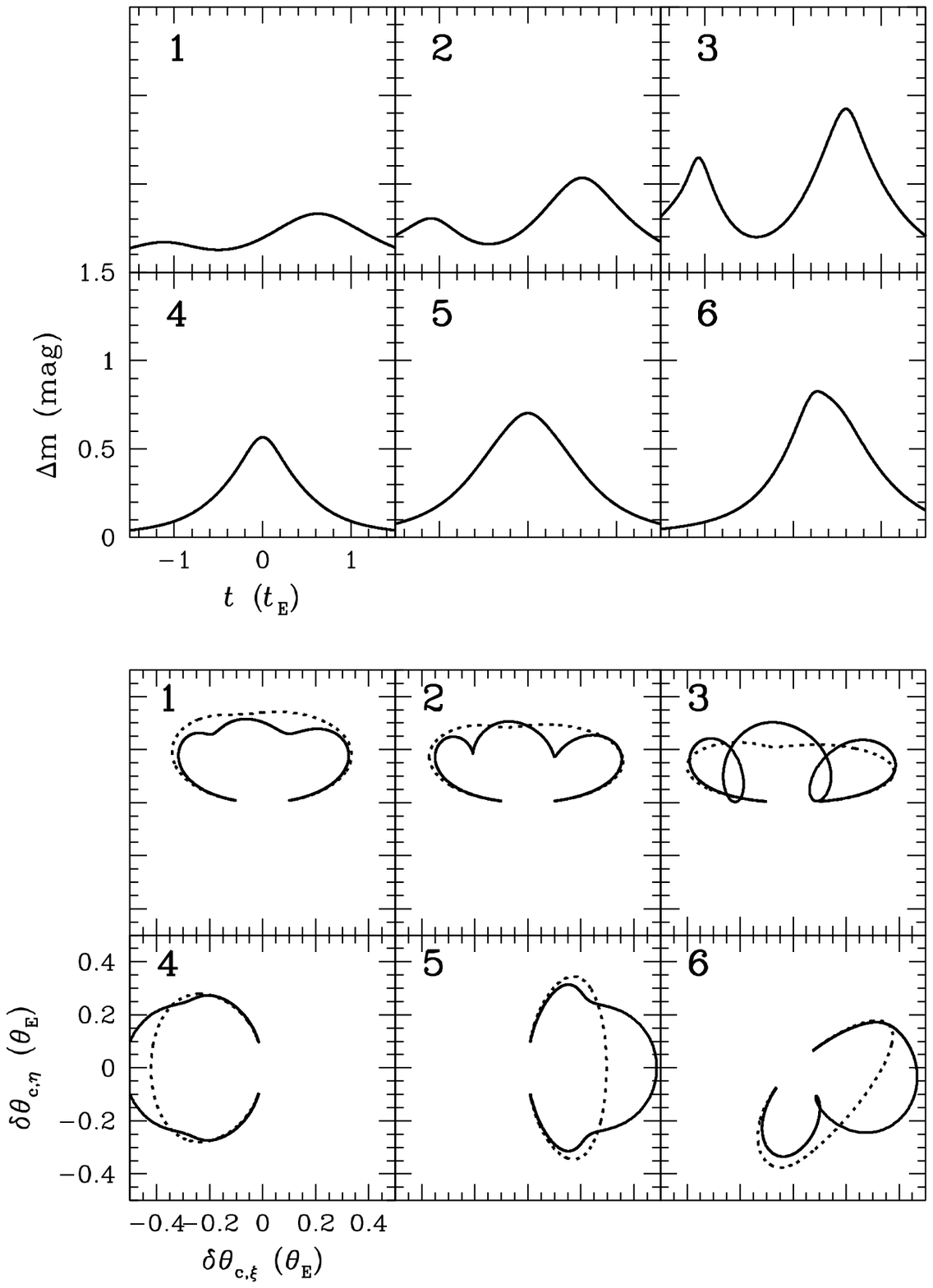}{1.15}
\noindent
{{\bf Figure 6:}\
Light curves (upper 6 panels) and astrometric  trajectories (lower 6 panels) 
for several example non-caustic crossing wide binary lensing events whose
source trajectories are marked in Fig.\ 4.  The notations are same 
as those in Fig.\ 3.
}

\postscript{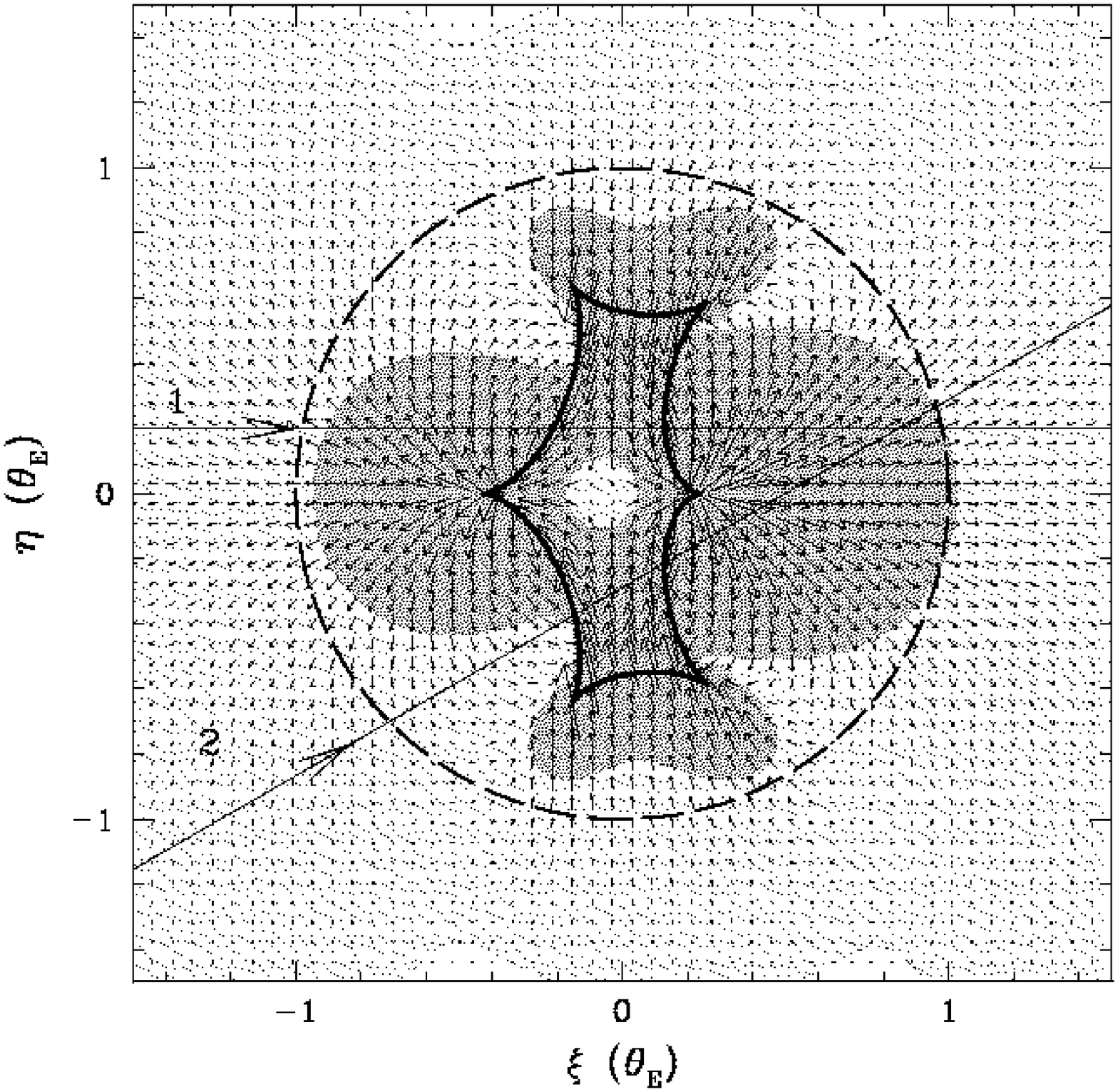}{1.0}
\noindent
{{\bf Figure 7:}\
Vector field map of the excess centroid shifts for a binary lens system with
$\ell=1.0$ (upper panel) and $q=0.5$.  The notations are same as those in 
Fig.\ 1.
}

\postscript{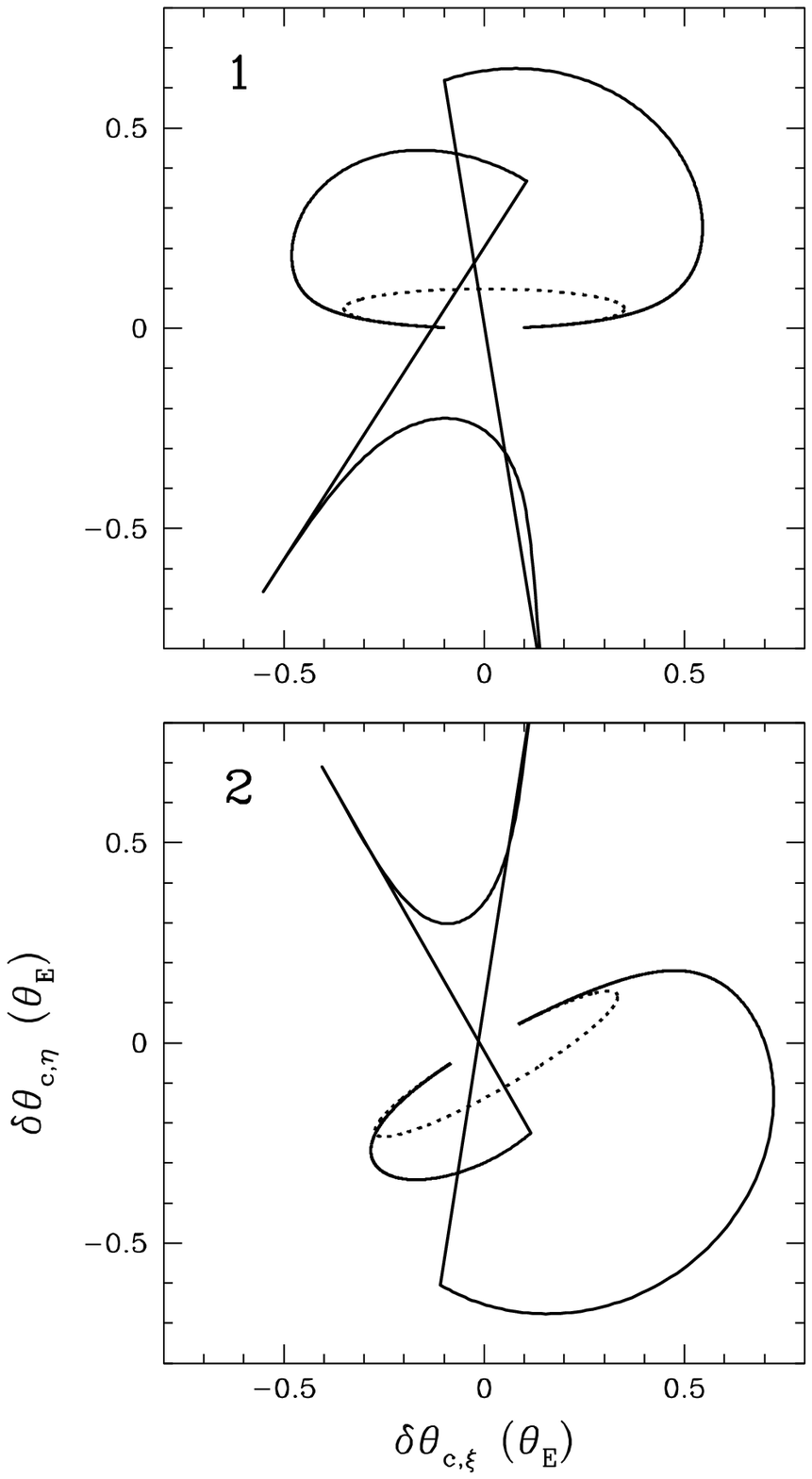}{1.2}
\noindent
{{\bf Figure 8:}\
The astrometric trajectories of caustic crossing events
whose source trajectories are marked in Fig\ 7.  
The dotted curves are the corresponding single lensing events.
}


\bigskip
\clearpage 


\begin{references}
\reference{}  Afonso, C.\ 1999, \aap, 344, 63
\reference{}  Alard, C., Mao, S., \& Guibert, J.\ 1995, \aap, 300, L17
\reference{}  Albrow, M., et al.\ 1996, in IAU Symp.\ 173, Astrophysical 
	      Applications of Gravitational Lensing ed.\ C.\ S.\ Kochanek 
	      \& J.\ N.\ Hewit (Dordrech: Kluwer), 227
\reference{}  Albrow, M., et al.\ 1999, \apj, 522, 1022
\reference{}  Alcock, C., et al.\ 1999a, \apj, 518, 44
\reference{}  Alcock, C., et al.\ 1999b, preprint (astro-ph/9907369)
\reference{}  Alcock, C., et al.\ 2000, preprint (astro-ph/0002510)
\reference{}  An, J., Han, C., \& Park, S.-H.\ 2000, in preparation
\reference{}  Bennett, D.\ P., et al.\ 1995, in AIP Conf.\ Proc.\ 336,
	      Dark matter, ed.\ Holt, S.\ S.\ and Bennett, C.\ L.
	      (New York: AIP), 77
\reference{}  Boden, A.\ F., Shao, M., \& Van Buren, D.\ 1998, \apj, 502, 538
\reference{}  Chang, K., \& Han, C.\ 1999, \apj, 525, 434
\reference{}  Colavita, M.\ M., et al.\ 1998, Proc.\ SPIE, 3350-31, 776
\reference{}  Di Stefano, R., \& Mao, S.\ 1999, \apj, 457, 93
\reference{}  Di Stefano, R., \& Perna, R.\ 1997, \apj, 488, 55
\reference{}  Di Stefano, R., \& Scalzo, R.\ J.\ 1999, \apj, 512, 579
\reference{}  Dominik, M.\ 1999, \aap, 341, 943
\reference{}  Dominik, M., \& Hirshfeld, A.\ C.\ 1996, \aap, 313, 841
\reference{}  Gaudi, B.\ S., \& Gould, A.\ 1997, \apj, 482, 83
\reference{}  Han, C.\ 1999, \mnras, 308, 1077
\reference{}  Han, C.\ 2000, \mnras, submitted
\reference{}  Han, C., \& Kim, T.-W.\ 1999, \mnras, 305, 795
\reference{}  H\o\hskip-1pt g, E., Novikov, I.\ D., \&
              Polnarev, A.\ G.\ 1995, \aap, 294, 287
\reference{}  Jeong, Y., Han, C., \& Park, S.-H.\ 1999, \apj, 511, 569
\reference{}  Mao, S., \& Di Stefano, R.\ 1995, \apj, 440, 22
\reference{}  Mao, S., \& Witt, H.\ J.\ 1998, \mnras, 300, 1041
\reference{}  Mariotti, J.\ M., et al.\ 1998, Proc.\ SPIE, 3350-33, 880
\reference{}  Miralda-Escud\'e, J.\ 1996, \apj, 470, L113
\reference{}  Miyamoto, M., \& Yoshii, Y.\ 1995, \aj, 110, 1427
\reference{}  Paczy\'nski, B.\ 1998, \apj, 404, L23
\reference{}  Pratt, M., et al.\ 1996, in IAU Symp.\ 173, Astrophysical 
              Applications of Gravitational Lensing ed.\ C.\ S.\ Kochanek
              \& J.\ N.\ Hewit (Dordrech: Kluwer), 221
\reference{}  Rhie, S.\ H., \& Bennett, D.\ P.\ 1996, Nucl.\ Phys.\
	      Proc.\ Suppl., 51B, 86
\reference{}  Udalski, A., Kubiak, M., \& Szyma\'nski, M.\ 1997,
	      Acta Astron., 47, 319
\reference{}  Udalski, A., Szyma\'nski, M., Mao, S., Di Stefano, R., 
	      Kaluzny, J., Kuniak, M., Mateo, M., \& Krzemi\'nski, W.\ 
	      1994, \apj, 436, 103
\reference{}  Unwin, S., Boden, A., \& Shao, M.\ 1997, in AIP Proc.\ 387, 
	      Space Technology and Applications International Forum 1997, 
	      ed.\ M\ S.\ El-Genk (New York: AIP), 63
\reference{}  Walker, M.\ A.\ 1995, \apj, 453, 37
\reference{}  Witt, H.\ 1990, \aap, 263, 311

\end{references}
\end{document}